\documentclass[showpacs]{revtex4}
\usepackage{graphicx}
\usepackage{amsmath}
\usepackage{amssymb}
\usepackage{color}
\usepackage[cp1251]{inputenc}
\usepackage[russian]{babel}

\begin{document}
\title{Usual and phantom scalar fields in five dimensions:\\
compactification and flat thick brane solutions}
\author{Vladimir Dzhunushaliev
\footnote{Senior Associate of the Abdus Salam ICTP}}
\email{vdzhunus@krsu.edu.kg} \affiliation{Dept. Phys. and Microel.
Engineer., Kyrgyz-Russian Slavic University, Bishkek, Kievskaya Str.
44, 720021, Kyrgyz Republic}

\author{Vladimir Folomeev}
\email{vfolomeev@mail.ru} \affiliation{Institute of Physics of NAS
KR, 265 a, Chui str., Bishkek, 720071,  Kyrgyz Republic}

\author{Shynaray Myrzakul and Ratbay Myrzakulov}
\email{cnlpmyra1954@yahoo.com, cnlpmyra@mail.ru}
\affiliation{Dept. Gen. and Theor. Phys., Eurasian National University, Astana, 010008, Kazakhstan} 

\begin{abstract}
In the model of a gravitating system with two scalar fields (one of which is phantom), two new types of regular solutions are found: mechanism for compactification of an extra dimension and a flat thick brane solution. It is shown that the first model has solutions oscillating over the extra coordinate and giving a finite radius of compactification of the fifth dimension and the second model is a flat thick brane embedded in the 5D Minkowski spacetime. Geometry of both models corresponds to a five-dimensional Minkowski space-time. Consideration of linear perturbations shows stability of the obtained solutions.
\end{abstract}

\maketitle

\section{Introduction}
\label{introd}

At the present time, there are two ways at consideration of extra dimensions in the multidimensional theories: the first  one assumes that the extra dimensions are small and compactified  by using some mechanism (compactification mechanism); the second one assumes that the extra dimensions have large or even infinite size, but we live on a thin four-dimensional sheet (brane) embedded into a higher-dimensional space called the bulk. Existence of the extra dimensions leads to the problem of their non-observability. This problem is being solved by using some compactification mechanisms in the first case, and by introducing various kinds of braneworld models in the second one. At the present time, apparently, there is no any unified mechanism working in both cases. In this paper we suggest such a unified model providing both
the compactification mechanism and existence of a brane-like solution.

The problem of compactification of extra dimensions is one of the most important problems in Kaluza-Klein theories. It is assumed that we live in a space with $d > 4$ dimensions, but that $(d-4)$ of these dimensions have been compactified. The radius of curvature of space in the corresponding directions is of order $M_{pl}^{-1}$. That is why we cannot move in those directions, and space is apparently four-dimensional.

For solution of this problem, a few mechanisms of the spontaneous compactification were  suggested. The main idea of all these mechanisms consists in a search of special (vacuum) solutions of the multidimensional Einstein equations corresponding to representation of $d-$dimensional manifold in the form $M^d=M^4\times B^{d-4}$, where $M^4$ is four-dimensional space-time (it is desirable that this would be a Minkowski space-time) and $B^{d-4}$ is a compact ``internal'' space.

There are the following mechanisms of the spontaneous compactification \cite{aref}:
\begin{enumerate}
	\item Freund-Rubin compactification \cite{fre} with special ansatz for antisymmetric tensors, and also the Englert compactification \cite{englert};
	\item compactification by setting a gauge field in an internal space equal to the spin connection, suitable embedded in a gauge group \cite{volkov};
	\item monopole or instanton mechanism \cite{ranj};
	\item compactification by using scalar chiral fields \cite{omero};
	\item compactification by using radiative corrections \cite{witt}.
\end{enumerate}

Here we suggest a new (in our opinion) and simple compactification mechanism of the fifth dimension based on using of usual and phantom scalar fields. Thereupon, energy-momentum tensors  of the fields have different signs. For example, energy densities of the usual and phantom fields will be positive and negative, respectively. It allows us to consider the models when the energy-momentum tensors of both fields \emph{compensate} each other exactly. Then the five-dimensional Einstein equations have solutions which can be represented as product of a four-dimensional Minkowski space and compactified fifth dimension. For this purpose, potential energy of the scalar fields (usual and phantom) is being chosen in such a way to obtain a solution oscillating over the fifth dimension. That is why one can consider the fifth coordinate as being twisted into a circle. Choosing a mass of scalar fields, one can get any desirable radius of compactification, including the Planck value.

In the case of braneworld models, there are two kinds of models: thin and thick branes. The thin brane is an infinitely thin sheet embedded into an external multidimensional space and having delta-like localization of matter~\cite{rs}. However, from the physical point of view, it is most reasonable to consider models of a thick brane \cite{Bronnikov:2005bg}-\cite{Dzhun2} having finite thickness in extra dimensions. Both models give asymptotically anti-de Sitter (or de Sitter) solutions far from the brane. In this paper we consider another type of thick brane solutions when the warp factor is a constant over the bulk. This case corresponds to a Minkowski space-time. One can call such type of solutions as {\it flat thick brane solutions}. One can consider these solutions as follows: in the simplest case the five-dimensional action for a brane can be presented in the following form
$$
S=-\frac{1}{16\pi G_{(5)}}\int \sqrt{g^{(5)}}\left(R^{(5)}-2\Lambda\right)d^{\,4} x\, dz-\sigma\int\sqrt{g^{(4)}}d^{\,4} x.
$$
Here $z$ is the fifth coordinate, 
$\Lambda$ is a five-dimensional cosmological constant, $\sigma$ can be regarded as the brane tension (the energy density
per unit three-dimensional volume). According to the Randall-Sundrum's second model~\cite{rs}, there is a fine tuning condition
between the brane and bulk parameters $\Lambda$ and $\sigma$:
$$
\Lambda=-\frac{4\pi}{3}G_{(5)}\sigma^2.
$$
For such a model, there are two kinds of asymptotic solutions: if $\sigma>0$, then one has the anti-de Sitter solution with the
warp factor $a(z)=\exp{\left(-k |z|\right)}$, where $k=(4\pi/3)G_{(5)}\sigma$. If $\sigma<0$, then one has the de Sitter solution.
In this paper we will consider a case when the brane tension $\sigma=0$. This is being achieved by using the usual and phantom scalar fields
with positive and negative energy densities compensating  each other. That is why this type of solutions can be regarded as 
 flat thick brane solutions. 

\section{General equations}

We start with the Lagrangian
\begin{equation}
\label{lagrangian}
    L =-\frac{R}{2}+
\epsilon_1 \left[\frac{1}{2}\partial_\mu \varphi \partial^\mu\varphi-V_1(\varphi)\right]+
\epsilon_2 \left[\frac{1}{2}\partial_\mu \chi \partial^\mu\chi-V_2(\chi)\right]~,
\end{equation}
where $R$ is the five-dimensional scalar curvature, $\varphi, \chi$ are two noninteracting scalar fields with the potentials  $V_1(\varphi)$ and $V_2(\chi)$, respectively. The constants $\epsilon_{1,2}=\pm 1$ correspond to either usual ($\epsilon=+1$) or phantom scalar field ($\epsilon=-1$). Taking the five-dimensional metric in the form
\begin{equation}
\label{metric}
ds^2=a^2(z)\left( \eta_{\mu\nu}dx^{\mu}dx^{\nu}-dz^2\right),
\end{equation}
where Greek indices $\mu,\nu...$ refer to four
dimensions, $\eta_{\mu\nu}=\{1,-1,-1,-1\}$, and the function $a(z)$ depends only on the fifth coordinate $z$.
The energy-momentum tensor can be obtained from \eqref{lagrangian}
\begin{equation}
\label{emt}
	T_B^A=\epsilon_1\partial_B \varphi \partial^A \varphi+
	\epsilon_2\partial_B \chi \partial^A \chi-
		\delta^A_B\left\{
		\epsilon_1 \left[
			\frac{1}{2}\partial_C \varphi \partial^C \varphi-V_1(\varphi)
		\right] +
		\epsilon_2\left[
			\frac{1}{2}\partial_C \chi \partial^C \chi - V_2(\chi)
		\right]
	\right\},
\end{equation}
where Latin indices
 $A, B, C...$ refer to five dimensions.

By varying \eqref{lagrangian} with use of the metric \eqref{metric}, one can obtain the Einstein equations
\begin{eqnarray}
\label{einst1}
G_\alpha^\alpha=-3\frac{a^{\prime\prime}}{a^3}=T_\alpha^\alpha,\\
G_z^z=-6\frac{a^{\prime 2}}{a^4}=T^z_z,
\label{einst2}
\end{eqnarray}
where the first equation corresponds to the four four-dimensional Einstein equations, and the second one is the fifth component
of the Einstein equations. The equations for scalar fields will be
\begin{eqnarray}
\label{sf}
\varphi^{\prime \prime}+3 \mathcal{H} \varphi^{\prime}=a^2 \frac{\partial V_1(\varphi)}{\partial \varphi},\\
\chi^{\prime \prime}+3 \mathcal{H} \chi^{\prime}=a^2 \frac{\partial V_2(\chi)}{\partial \chi},
\end{eqnarray}
where $\mathcal{H}=a^{\prime}/a$.

Let us consider the simplest case when
$\epsilon_1=+1$, $\epsilon_2=-1$ and $\varphi=\chi$, $V_1(\varphi)=V_2(\chi)$.
Then one can see from
\eqref{emt} that $T_\alpha^\alpha=T_z^z=0$. In  this case, it follows from \eqref{einst1}-\eqref{einst2} that
\begin{equation}
\label{metf}
    a=const.
\end{equation}
The coordinates
$x^\mu$ can be chosen in such a way that $a=1$.

\subsection{Compactification of the fifth dimension}

Choosing the potential energy of scalar field in the simplest form
$$
V(\varphi)=-\frac{1}{2}m^2\varphi^2,
$$
where $m$ is a mass of the scalar field, it follows
from \eqref{sf} that
\begin{equation}
\label{back_sol_c}
\varphi=\chi=\varphi_0 \cos m z,
\end{equation}
where $\varphi_0$ is an integration constant. Eq. \eqref{back_sol_c} means that the functions $\varphi(z)$ and $\chi(z)$ are periodical with the period $\lambda=2\pi/m$. So one can say that the fifth dimension is a circle $S^1$ with some compactification radius $R=n/m$, where $n=1,2,3...$ The coordinate $z$ runs from $0$ to $2\pi R$, and the points $z=0$ and $z=2\pi R$ are identified.

\subsection{Flat thick brane solution}

Choosing the potential energy of scalar field in the form
$$
V(\varphi)=\frac{1}{4}\lambda \left(\varphi^2-\frac{m^2}{\lambda}\right)^2,
$$
where $m$ is a mass of the scalar field and $\lambda$ is a self-coupling constant, one can obtain from \eqref{sf}
the following kink-like solution
\begin{equation}
\label{back_sol_w}
\varphi=\chi=\pm \frac{m}{\sqrt{\lambda}}\tanh{\left[\frac{m}{\sqrt{2}}(z-z_0)\right]},
\end{equation}
where $z_0$ is an integration constant. In order to have a brane-like solution at $z=0$, we have to choose $z_0 = 0$.
I.e., we have obtained the five-dimensional kink-like solution with the flat Minkowski space-time. This solution can be
interpreted as a flat  brane solution. And this is the distinction between  usual thick brane solutions with asymptotically
anti-de Sitter (or de Sitter) space and the solution obtained here.

\section{Stability analysis}

Let us use the method suggested in \cite{Kobayashi:2001jd} for stability analysis of the obtained solution. For this purpose, we rewrite the metric \eqref{metric} in the perturbed form
\begin{equation}
\label{metric_pert}
ds^2=a^2(z)\left\{\left[1+2\psi(x^\mu,z)\right]\eta_{\mu\nu}dx^{\mu}dx^{\nu}-\left[1+2\phi(x^\mu,z)\right]dz^2\right\},
\end{equation}
where $\psi(x^\mu,z), \phi(x^\mu,z)$ are perturbations of the metric. Similarly, let us search for perturbed solutions for the scalar fields as
$$
\varphi=\varphi_b(z)+\delta\varphi(x^\mu,z),\quad \chi=\chi_b(z)+\delta\chi(x^\mu,z),
$$
where $\varphi_b(z), \chi_b(z)$ are the background solutions \eqref{back_sol_c} and  \eqref{back_sol_w}. Then equations for the perturbations will be
\begin{eqnarray}
\label{perturb1}
&&(z,z): 3\eta^{\rho\lambda}\psi_{;\rho\lambda}+12\mathcal{H}\psi^{\prime}-12\mathcal{H}^2\phi=\varphi_b^\prime\delta\theta^\prime
-a^2\frac{\partial V(\varphi)}{\partial \varphi_b}\delta\theta,\\
\label{perturb2}
    &&(z,\mu): -3\psi^\prime_{;\mu}+3\mathcal{H}\phi_{;\mu}=\varphi_b^\prime\delta\theta_{;\mu},\\
    \label{perturb3}
    &&(\mu,\nu): \left(3\psi^{\prime\prime}-6\mathcal{H}^\prime\phi-3\mathcal{H}\phi^\prime+
    9\mathcal{H}\psi^\prime-6\mathcal{H}^2\phi+\eta^{\rho\lambda}\phi_{;\rho\lambda}
    +2\eta^{\rho\lambda}\psi_{;\rho\lambda}\psi\right)\delta^\mu_\nu \\ \nonumber
    &&\qquad \qquad-\eta^{\rho\lambda}\phi_{;\rho\lambda}-2\eta^{\rho\lambda}\psi_{;\rho\lambda}=
    \left(-\varphi_b^\prime\delta\theta^\prime-a^2\frac{\partial V(\varphi)}{\partial \varphi_b}\delta\theta\right),\\
    \label{perturb4}
    &&\text{matter}: \delta\theta^{\prime\prime}+3\mathcal{H}\delta\theta^\prime+\eta^{\rho\lambda}\theta_{;\rho\lambda}=
    a^2\frac{\partial^2 V(\varphi)}{\partial \varphi_b^2}\delta\theta,
\end{eqnarray}
where the notation $\delta\theta=\delta\varphi-\delta\chi$ is introduced, and the semicolon denotes the covariant derivative with respect to $\eta_{\mu\nu}$. From the off-diagonal part of \eqref{perturb3}, we have that
\begin{equation}
\label{rel}
\phi+2\psi=0.
\end{equation}
Taking into account \eqref{rel}, one can obtain from \eqref{perturb2}
\begin{equation}
\label{dt}
\delta\theta=-\frac{3}{\varphi^\prime_b}\left(\psi^\prime+2\mathcal{H}\psi\right).
\end{equation}

By subtracting \eqref{perturb1} from \eqref{perturb3} and taking into account that
$$
\mathcal{H}^\prime=\mathcal{H}^2
$$
(it follows from \eqref{einst1}-\eqref{einst2}), we will have the following master equation of the system
\begin{equation}
\label{master}
    \psi^{\prime\prime}+\eta^{\rho\lambda}\psi_{|\rho\lambda}
    +\left(3\mathcal{H}
    -2\frac{\varphi_b^{\prime\prime}}{\varphi_b^{\prime}}\right)\psi^{\prime}
    +4\left(\mathcal{H}^{\prime}-\mathcal{H}
    \frac{\varphi_b^{\prime\prime}}{\varphi_b^{\prime}}
    \right)
    \psi=0.
\end{equation}

In order to examine the stability of the system, let us transform this equation into the Shr\"{o}dinger-like equation. For this purpose, let us introduce the new function $F(z,x^{\mu})$
\begin{equation}
    \psi(z,x^{\mu})= \varphi_b^{\prime}(z) F(z,x^{\mu}).
\end{equation}
Then we will have the following equation for the scalar perturbation
\begin{equation}
-F^{\prime\prime}(z, x^{\mu})+ V_{e}(z)\cdot F(z, x^{\mu}) =
\eta^{\rho\lambda}F(z, x^{\mu})_{;\rho\lambda},
\label{Schrodinger equation}
\end{equation}
where the effective potential takes the form
\begin{equation}
V_{e} = -\frac{5}{2}\mathcal{H}^{\prime}
+\frac{9}{4}\mathcal{H}^2
+\mathcal{H}\frac{\varphi_b^{\prime\prime}}{\varphi_b^{\prime}}
-\frac{\varphi_b^{\prime\prime\prime}}{\varphi_b^{\prime}}
+2\left(\frac{\varphi_b^{\prime\prime}}{\varphi_b^{\prime}}\right)^2
.
\label{effective potential}
\end{equation}

In our case, taking into account the solutions \eqref{metf} and \eqref{back_sol_c}, the effective potential for the case of compactification becomes
$$
V_e=-m^2+2 m^2 \cot^2(m z)
,
$$
and for the flat thick brane case, using \eqref{metf} and \eqref{back_sol_w}, we have
$$
V_e=m^2\left[1+\tanh^2{\left(\frac{m}{\sqrt{2}}(z-z_0)\right)}\right]
.
$$
Let us expand the Shr\"{o}dinger equation \eqref{Schrodinger equation} into Fourier integral in a Minkowski space as follows
\begin{equation}
F(z,x^{\mu}) = \int \frac{d^4 p}{(\sqrt{2\pi})^4}
f(z) e^{ip_{\mu}x^{\mu}}.
\label{expansion for flat}
\end{equation}
Then we will have
\begin{equation}
\label{four_mod}
-f^{\prime \prime}(z)+V_e(z)f(z)=\mu^2 f(z)\,,
\end{equation}
where $\mu^2=-p^2$, $p$ is the momentum from \eqref{expansion for flat}. For stable solutions, it is necessary that
$\mu^2>0$.

Let us search for a solution of the last equation as follows:

\subsection{The case of compactification}

It is convenient to rewrite Eq.~\eqref{four_mod} in the form
$$
-f^{\prime \prime}-\left[E-V_0 \cot^2(m z)\right]f=0,
$$
where $E=\mu^2+m^2$, $V_0=2 m^2$, $m=\pi/b$, $b$ is the width of the potential well. Energy levels of this equation, satisfying the condition $f=0$ on the bounds of the potential well at $z=0$ and $z=b$, are
$$
E_n=(n^2+2n-1)m^2,\quad n=1,2,3...
$$
The value $E_1=2 m^2$ corresponds to the ground level, in what follows that
$\mu^2=m^2>0$, i.e., the solution is stable.

The normalized wave function of the ground level $n=1$ is
$$
f=\frac{2}{\sqrt{3}}\sin^2\left(\frac{\pi z}{b}\right).
$$
This solution can be being considered as a periodical solution with the same period as in \eqref{back_sol_c}.

\subsection{The case of the flat thick brane}

One can rewrite Eq.~\eqref{four_mod} in the form
\begin{equation}
	f^{\prime\prime}(x) + 2\left[ E-V(x) \right] f(x) = 0,
\end{equation}
where $x=m(z-z_0)/\sqrt{2}$, $E=\left[(\mu/m)^2-1\right]$ and $V(x)=\tanh^2(x)$ is always positive. Then it is obvious that the
energy levels $E$ will be always positive  as well. That is why $\mu^2>0$, i.e., the system is stable if a discrete spectrum of $E$ does exist.

To clarify this,
let us find energy spectrum for the case under consideration. Introducing a new variable $\xi=\tanh{x}$, the last equation gives
\begin{equation}
	\frac{d^2 f}{d\xi^2}-2\xi\frac{d f}{d\xi} +
	2\left( E-\xi^2 \right) f = 0
\end{equation}
with the solution
\begin{equation}
	f=C_1\exp{\left(\frac{1-\sqrt{3}}{2}\,\xi^2\right)}
	F_1\left[\frac{-1+\sqrt{3}-2E}{4\sqrt{3}},\frac{1}{2},\sqrt{3}\,\xi^2\right],
\end{equation}
where $F_1$ is the Kummer confluent hypergeometric function.
This function is finite at $\xi=\pm 1$ (i.e., at $x=\pm \infty$) if
\begin{equation}
	\frac{-1+\sqrt{3}-2E}{4\sqrt{3}}=-n,
\end{equation}
where
$n=0, 1, 2, ...$ (then $F$ is a finite polynomial of the power $n$ at $\xi=\pm 1$). Using this condition, we will have the following energy levels
\begin{equation}
	E_n=\frac{\sqrt{3}(4n+1)-1}{2}.
\end{equation}
Taking into account that the depth of the potential well  $V(x)=\tanh^2(x)$ is equal to 1,
one can easily see that there exists only one discrete energy level at $n=0$ (the ground level), i.e., the system is stable. For other $n$, one has
a continuous spectrum.

\section{Conclusions}

We have considered a possibility of using the model with two non-interacting scalar fields~-- usual and phantom ones~-- in application to the five-dimensional problems.
There were suggested two models: the new compactification mechanism of the extra dimension and the flat thick brane solution.
In  both models, the potential energy of scalar fields is being chosen identical. In the first model, we have used the simplest quadratic form of the potential energy. It allows us to find
 solutions which can be being considered as solutions with the compactified fifth dimension (a circle $S^1$). In the second model, we have used the ``Mexican hat'' potential energy. The obtained solution corresponds to the flat thick brane solution in the five-dimensional space-time. In both cases, it was shown that the solutions are stable against linear perturbations.

The described mechanism of compactification of one extra dimension can be generalized for a greater number of extra dimensions.


\begin{thebibliography}{0}

\bibitem{aref}
I.Y. Arefeva, I.V. Volovich, Sov. Phys. Usp. {\bf 28}, 694 (1985).

\bibitem{fre}
P.G.O. Freund, M.A. Rubin, Phys. Lett.  {\bf B 94}, 233 (1980).

\bibitem{englert}
F. Englert, Ibidem.  {\bf 119}, 339 (1982).

\bibitem{volkov}
D. V. Volkov, V.I. Tkach, Pis'ma Zh. Eksp. Teor. Fiz. {\bf 32}, 681 (1980); Teor. Mat. Fiz. {\bf 51}, 171 (1982);\\
S. Randjbar-Daemi, R. Percacci, Phys. Lett.  {\bf B 117}, 41 (1982).

\bibitem{ranj}
S. Randjbar-Daemi, A. Salam, J. Strathdee, Nucl. Phys.  {\bf B 242},
447 (1984).

\bibitem{omero}
C. Omero, R. Percacci, Ibidem.  {\bf 165}, 351 (1980);\\
M. Gell-Mann, B. Zwiebach, Phys. Lett.  {\bf B 141}, 333 (1984).

\bibitem{witt}
E. Witten, Nucl. Phys.  {\bf B 195}, 481 (1982);\\
P. Candelas, S. Weinberg, Nucl. Phys.  {\bf B 237}, 397 (1984).

\bibitem{rs}
  L.~Randall and R.~Sundrum,
  Phys.\ Rev.\ Lett.\  {\bf 83}, 3370 (1999)
  [arXiv:hep-ph/9905221];\\
 L.~Randall and R.~Sundrum,
  Phys.\ Rev.\ Lett.\  {\bf 83}, 3690 (1999)
  [arXiv:hep-th/9906064].

\bibitem{Bronnikov:2005bg}
K.~A.~Bronnikov and B.~E.~Meierovich,
J.\ Exp.\ Theor.\ Phys.\  {\bf 101}, 1036 (2005), [Zh.\ Eksp.\
Teor.\ Fiz.\  {\bf 101}, 1184 (2005)], gr-qc/0507032.

\bibitem{Barbosa-Cendejas:2005kn}
N.~Barbosa-Cendejas and A.~Herrera-Aguilar,
JHEP {\bf 0510}, 101 (2005), hep-th/0511050.

\bibitem{DeWolfe:1999cp}
O.~DeWolfe, D.~Z.~Freedman, S.~S.~Gubser and A.~Karch,
Phys.\ Rev.\ D {\bf 62}, 046008 (2000), hep-th/9909134.

\bibitem{Bazeia:2004dh}
D.~Bazeia and A.~R.~Gomes,
JHEP {\bf 0405}, 012 (2004), hep-th/0403141.

\bibitem{Dzhun}
V. Dzhunushaliev, ``Thick brane solution in the presence of two
interacting scalar fields'', gr-qc/0603020.

\bibitem{Dzhun1}
 V.~Dzhunushaliev, V.~Folomeev, D.~Singleton and S.~Aguilar-Rudametkin,
  Phys.\ Rev.\  D {\bf 77}, 044006 (2008),
  hep-th/0703043.

\bibitem{Dzhun2}
V. Dzhunushaliev, V. Folomeev, K. Myrzakulov, R. Myrzakulov,
``Thick brane in 7D and 8D spacetimes'', gr-qc/0705.4014.


\bibitem{Kobayashi:2001jd}
  S.~Kobayashi, K.~Koyama and J.~Soda,
  Phys.\ Rev.\  D {\bf 65}, 064014 (2002)
  [arXiv:hep-th/0107025].



\end{thebibliography}
\end{document}